\documentclass[prl,showpacs,preprintnumbers,amsmath,amssymb,twocolumn,superscriptaddress]{revtex4}

\usepackage{graphicx}
\usepackage{dcolumn}
\usepackage{bm}
\usepackage{color}

\usepackage{amsmath}	
\usepackage{amsfonts}

\newcommand{\calL}{{\mathcal{L}}}
\newcommand{\calLz}{{\mathcal{L}_z}}

\newcommand{\vecc}[1]{\mbox{\boldmath $#1$}}
\newcommand{\Lztot}{L_z}

\begin{document}

\title{
Orbital Angular Momentum and Spectral Flow
in Two Dimensional Chiral Superfluids
}

\author{Yasuhiro Tada}
\affiliation{Institute for Solid State Physics, The University of
  Tokyo, Kashiwa 277-8581, Japan}
\author{Wenxing Nie}
\thanks{Corresponding author: {\tt wenxing.nie@gmail.com}}
\affiliation{Institute for Advanced Study, Tsinghua University,
Beijing 100084, China}
\affiliation{Institute for Solid State Physics, The University of
  Tokyo, Kashiwa 277-8581, Japan}
\author{Masaki Oshikawa}
\affiliation{Institute for Solid State Physics, The University of
  Tokyo, Kashiwa 277-8581, Japan}

\begin{abstract}
We study the orbital
angular momentum (OAM) $L_z$ in two dimensional 
chiral
$(p_x+ip_y)^{\nu}$-wave superfluids (SF)
of $N$ fermions on a disc at zero temperature, in terms of 
spectral asymmetry and spectral flow. 
It is shown that $L_z=\nu N/2$ for any integer $\nu$, in the BEC regime.
In contrast, in the BCS limit, 
while the OAM is $L_z=N/2$ for the $p+ip$-wave SF,
for chiral SF with $\nu\geq2$, 
the OAM is remarkably suppressed as
$L_z=N\times O(\Delta_0/\varepsilon_F)\ll N$,
where $\Delta_0$ is the gap amplitude and $\varepsilon_F$
is the Fermi energy.
We demonstrate that
the difference between the $p+ip$-wave SF and the other chiral SFs 
in the BCS regimes originates from the 
nature of edge modes and related depairing effects.
\end{abstract}



\maketitle


The orbital angular momentum (OAM) $\Lztot$ of chiral superfluids (SF)
of fermions
is a fundamental problem which has been under intense investigation
over several decades
\cite{book:Vollhart1990,
pap:Ishikawa1977,pap:McClure1979,
pap:Mermin1980,pap:Volovik1995,
book:Volovik2003,book:Leggett2006,pap:Kita1998,pap:Goryo1998,
pap:Stone2008,pap:Sauls2011,pap:Mizushima2008,
pap:Tsutsumi2012PRB,pap:Bradlyn2012,pap:Hoyos2013,pap:Shitade2014,
pap:Anderson1961,pap:Leggett1975,
pap:Volovik1975,pap:Cross1975,pap:Balatsky1985}.
Among the chiral SFs, the simplest $p+ip$-wave 
pairing state describes
the A-phase of liquid $^3$He~\cite{book:Vollhart1990}, 
and quite likely also the
superconducting phase of Sr$_2$RuO$_4$~\cite{pap:214_2003, pap:214_2012}.
The OAM is a direct manifestation of the
broken chiral symmetry~\cite{pap:Ikegami2013},
and is also closely related to
the edge current, which has been so far not observed experimentally
in Sr$_2$RuO$_4$ despite the expectation from the $p+ip$-wave
SF picture~\cite{pap:SQUID2010}.

Most of the existing studies on the OAM have focused on
$p+ip$-wave SF.
However, higher-order chiral SFs such as $d+id$, $f+if$, \ldots-wave
ones are also of interest~\cite{pap:Hasan2010,pap:Qi2011}, 
and have potential applications to
candidates for chiral superconductors, 
such as UPt$_3$~\cite{pap:UPt3_2002}, 
URu$_2$Si$_2$~\cite{pap:U122_2007}, and SrPtAs~\cite{pap:SrPtAs2011}.
We focus on fundamental chiral SFs with the pairing symmetry
$\sim(p_x+ip_y)^{\nu}$ which can be classified
by the integer angular momentum $\nu$
of each Cooper pair: $\nu=1$ corresponds to $p+ip$, $\nu=2$ to
$d+id$, and so on.
In fact, as we will demonstrate in this Letter, there is an unexpected
fundamental difference between the $p+ip$-wave and higher-order chiral
SFs with respect to the
OAM; thus it is essential to consider the
higher-order ones as well, for a complete understanding of the problem.

The main issue with the OAM is that
different viewpoints lead to different predictions for it,
resulting in an apparent paradox~\cite{pap:Leggett1975}.
One argument is that, since each Cooper pair has the OAM
$\nu$, $\Lztot = \nu N/2$
where $N$ is the total number of fermions.
There is, however, a different view starting from the
normal (non-superconducting) Fermi liquid, which has $\Lztot=0$.
Formation of Cooper pairs with an angular momentum
would change the value of $\Lztot$ 
from zero to a non-vanishing value in the chiral SF phase.
However, since only the low-energy fermions
near the original Fermi surface would be affected,
$\Lztot$ should be suppressed as
$(\Delta_0/\varepsilon_F)^{\gamma}N/2$ with $\gamma>0$,
where $\Delta_0$ is the pairing gap amplitude and
$\varepsilon_F$ is the Fermi energy.

Of course, the analysis did not stop at the hand-waving arguments
and many calculations have been carried out based on
various schemes, leading to different results.
We note that, in the limit of strong pairing of fermions,
the superfluid phase may be understood as a result of Bose-Einstein
condensation (BEC) of bosonic molecules.
In this limit, it would be natural to expect
that $\Lztot = \nu N/2$,
since each bosonic molecule carries the OAM $\nu$.
However, this does not necessarily imply that the same value
of $\Lztot$ persists in the regime where the superfluid is
described by Bardeen-Cooper-Schrieffer (BCS) theory.
In fact, the ``weak-pairing'' chiral SFs in the BCS regime
is a topological superfluid with gapless edge states,
while the ``strong-pairing'' chiral SFs in the BEC regime
is a non-topological one~\cite{pap:Read2000}.
Thus they are distinct superfluid phases separated by
a quantum phase transition, and thus $\Lztot$ could take
very different values. 
Even within the same phase,
the stability of the OAM has not been established.

Historically, strong reduction of $\Lztot$ ($\gamma \geq 1$) was
predicted for the BCS regime of $p+ip$-wave SF,
in several of
the earlier papers on the subject~\cite{pap:Anderson1961,
pap:Leggett1975,pap:Volovik1975,pap:Cross1975,
pap:Balatsky1985}. 
On the other hand, many others, including most of more recent
ones support the full OAM $\Lztot=N/2$ at zero
temperature, even for the BCS regime
\cite{pap:Ishikawa1977,pap:McClure1979,
pap:Mermin1980,pap:Volovik1995,
book:Volovik2003,book:Leggett2006,pap:Kita1998,pap:Goryo1998,
pap:Stone2008,pap:Sauls2011,pap:Mizushima2008,
pap:Tsutsumi2012PRB,pap:Bradlyn2012,pap:Hoyos2013,pap:Shitade2014}.
However, so far there is no clear physical picture
why $\Lztot=N/2$ holds even in the BCS regime.
Experimental investigation of the problem is difficult and
there have been very few reports so
far~\cite{pap:experiment1996,Ishikawa-IAM-expt-QFS}.
Therefore the long-standing paradox is not yet resolved
even for the $p+ip$-wave SF, let alone for the higher-order
ones with $\nu \geq 2$.


In this study, we investigate the problem in the simplest ideal setting:
two dimensional (2D) chiral SFs confined on a completely circular
disc with a specular wall at zero temperature, in the framework
of Bogoliubov-de Gennes (BdG) Hamiltonian.
For simplicity, we assume that the $d$-vector
is $\vecc{d}=(0,0,d_z)$ for 
the triplet states so that both the singlet states and the 
triplet states can be discussed in a parallel way; 
our analysis is also applicable
to the spinless fermions with slight modifications.
We consider the Hamiltonian
$\hat{H}=\int d^2x\psi^{\dagger}_{\sigma}[(p_x^2+p_y^2)/2m_0+V-\mu]
\psi_{\sigma}
+\int d^2x\psi_{\uparrow}^{\dagger}\Delta(r)(p_x+ip_y)^{\nu}
\psi_{\downarrow}^{\dagger}+({\rm h.c.}),$
where $p_{j}=-i\partial/\partial x_j$,
$m_0$ is the fermion mass, and $\mu$ is the chemical potential.
$V(r)$ describes the wall of the container, and is chosen to be
$V(r<R)=0$ and $V(r>R)=\infty$ with a radius $R$.
There is no texture in this model.

Although we only consider constant pair potentials
in numerical calculations,
our discussion is also applicable to systems with $\Delta(r)$
for which $r$-dependence is determined through self-consistent 
calculations.
The field operator is expanded in terms of a single particle 
basis as
$\psi_{\sigma}(\vecc{r})=\sum_{nl}c_{nl\sigma}\varphi_{nl}(\vecc{r})$
where $\varphi$ satisfies
$[(p_x^2+p_y^2)/2m_0+V(r)-\mu]\varphi_{nl}=\varepsilon_{nl}
\varphi_{nl}$
\cite{pap:Kita1998,pap:Stone2008}.
Then the Hamiltonian becomes
\begin{align}
\hat{H}&=\sum_{l}\sum_{nn^{\prime}}
\left[
\begin{array}{c}
c_{n,l+\nu,\uparrow}^{\dagger}\\ c_{n,-l,\downarrow}
\end{array}\right]^T\notag \\
&\quad \times \left[
\begin{array}{cc}
\varepsilon_{n,l+\nu}\delta_{nn^{\prime}} & \Delta_{nn^{\prime}}^{(l)}\\
\Delta_{n^{\prime}n}^{(l)\ast}& -\varepsilon_{n,-l}\delta_{nn^{\prime}}
\end{array}\right]
\left[
\begin{array}{c}
c_{n^{\prime},l+\nu,\uparrow}\\ c_{n^{\prime},-l,\downarrow}^{\dagger}
\end{array}\right],
\label{eq:Ham}
\end{align}
where $\Delta_{nn^{\prime}}^{(l)}=\int \varphi_{n,l+\nu}^{\ast}
\Delta (p_x+ip_y)^{\nu}\varphi_{n',-l}^{\ast}$,
with an appropriate high-energy regularization.
We denote the above matrix as $(H_{\rm BdG}^{(l)})_{nn^{\prime}}$.
The particle-hole symmetry connects different $l$-sectors as
$PH^{(l)}_{\rm BdG}P^{-1}= -H^{(-l-\nu)\ast}_{\rm BdG}$ where
$P=\sigma_x(i\sigma_y)$ in the Nambu space for odd (even) $\nu$,
which implies that, although eigenvalues come in pairs, each of them
lies in different $l$-sectors.

The OAM $\Lztot$ corresponds to the operator
$\hat{L}_z=\int d^2x \; \psi_{\sigma}^{\dagger}
\left(xp_y-yp_x\right)\psi_{\sigma}$, while
the total particle number operator is given as
$\hat{N}=\int d^2x \; \psi_{\sigma}^{\dagger}\psi_{\sigma}$.
These operators are clearly defined for the present model,
and include all the possible contributions.
Neither $\hat{L}_z$ nor
$\hat{N}$ commutes with the BdG
Hamiltonian~\eqref{eq:Ham} owing to the pairing term
in the Hamiltonian, and is not conserved.
Nevertheless, as pointed out in
Refs.~\cite{pap:Volovik1995,book:Volovik2003},
the combination
\begin{align}
\hat{{\mathcal L}}_z & \equiv \hat{L}_z - \frac{\nu}{2} \hat{N}
=  \sum_{nl\sigma} \left( l- \frac{\nu}{2} \right)
c_{nl\sigma}^{\dagger}c_{nl\sigma},
\end{align}
commutes with the Hamiltonian~\eqref{eq:Ham}, and thus is
a conserved quantity.
Physically, $\hat{\mathcal{L}}_z$ represents the correction
to the OAM with respect to its
``full'' value $\nu N/2$.
If the ground state belongs to the zero eigenvalue
sector of $\hat{\mathcal{L}}_z=0$, it follows that
$\Lztot=\nu N/2$.
However, $\hat{\mathcal{L}}_z$ could take different eigenvalues
in the ground state, as it is clear by considering
the limit of $\Delta \rightarrow 0$, where
$L_z=0$ and $\mathcal{L}_z=-\nu N/2$ hold.

Thus the eigenvalue $\mathcal{L}_z$ of $\hat{\mathcal{L}}_z$ in the
ground state is a nontrivial quantity.
In fact, it can still be calculated
exactly for the Hamiltonian~\eqref{eq:Ham}.
After the Bogoliubov transformation,
the ground state $|{\rm GS}\rangle$ is
simply the vacuum with respect to all the positive energy quasiparticles.
The eigenvalue 
${\mathcal L}_z$ for $|{\rm GS}\rangle$
can be obtained explicitly as
\begin{align}
{\mathcal L}_z&=
-\frac{1}{2}\sum_l\left(l+ \frac{\nu}{2}\right)\eta_l
,
& 
\eta_l&=\sum_m {\rm sgn}E^{(l)}_m,
\label{eq:L}
\end{align}
where $\{E^{(l)}_m\}_{m \in \mathbb{N}}$
are eigenvalues of $H^{(l)}_{\rm BdG}$, and
$\eta_l$ is called the spectral asymmetry
\cite{book:Volovik2003,pap:Paranjape1985,pap:NiemiSemenoff1986,pap:Stone1987}.
In actual calculations, we introduce a cut-off $M \gg 1$ so
that indices $n,n'$ are restricted not to exceed $M$.
Then $H_{\rm BdG}^{(l)}$ is represented by a $2M\times 2M$
matrix, and $\eta_l$ takes only even integer values.
We have verified that the results are independent
of $M$, when $M$ is sufficiently large.
From this formula, it is clear that
${\mathcal L}_z$ can change
only when there is a spectral flow, namely
some of the eigenvalues of $H_{\rm BdG}^{(l)}$ cross zero
as model parameters are varied.

\begin{figure}[htbp]
\begin{tabular}{lr}
\begin{minipage}{0.35\hsize}
\begin{center}
\includegraphics[width=\hsize,height=\hsize]{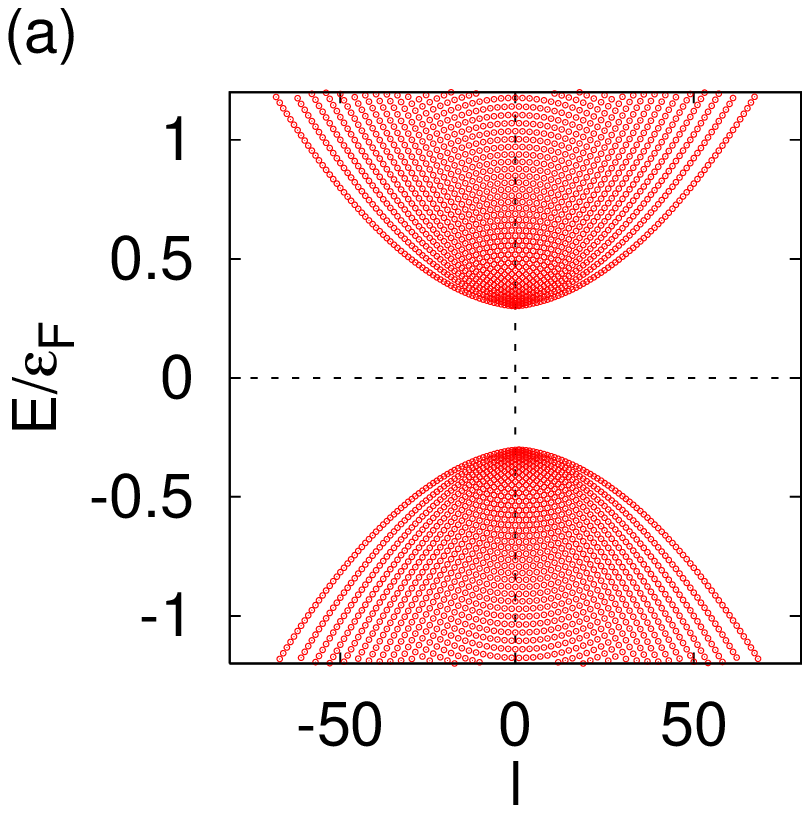}
\end{center}
\end{minipage}
\hspace{0.5cm}
\begin{minipage}{0.35\hsize}
\begin{center}
\includegraphics[width=\hsize,height=\hsize]{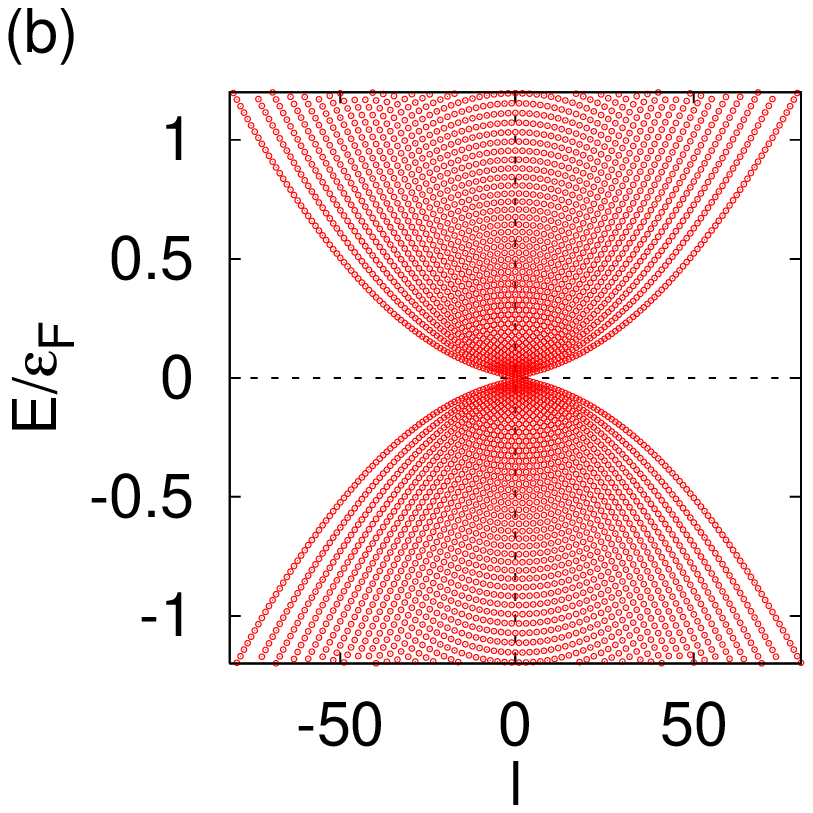}
\end{center}
\end{minipage}
\end{tabular}
\begin{tabular}{cc}
\begin{minipage}{0.35\hsize}
\begin{center}
\includegraphics[width=\hsize,height=\hsize]{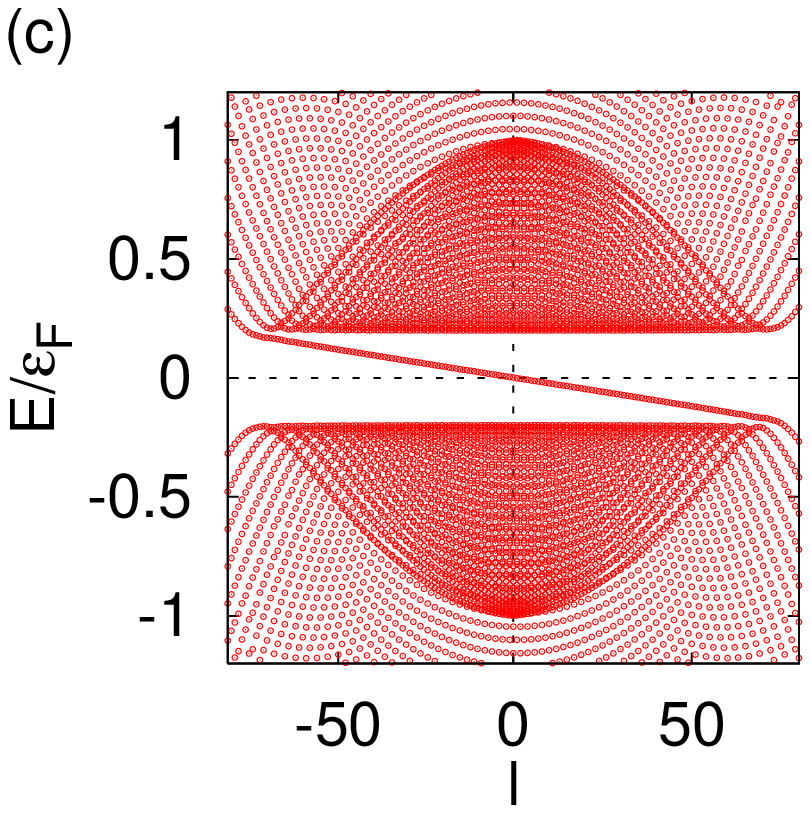}
\end{center}
\end{minipage}
\hspace{0.5cm}
\begin{minipage}{0.35\hsize}
\begin{center}
\includegraphics[width=\hsize,height=\hsize]{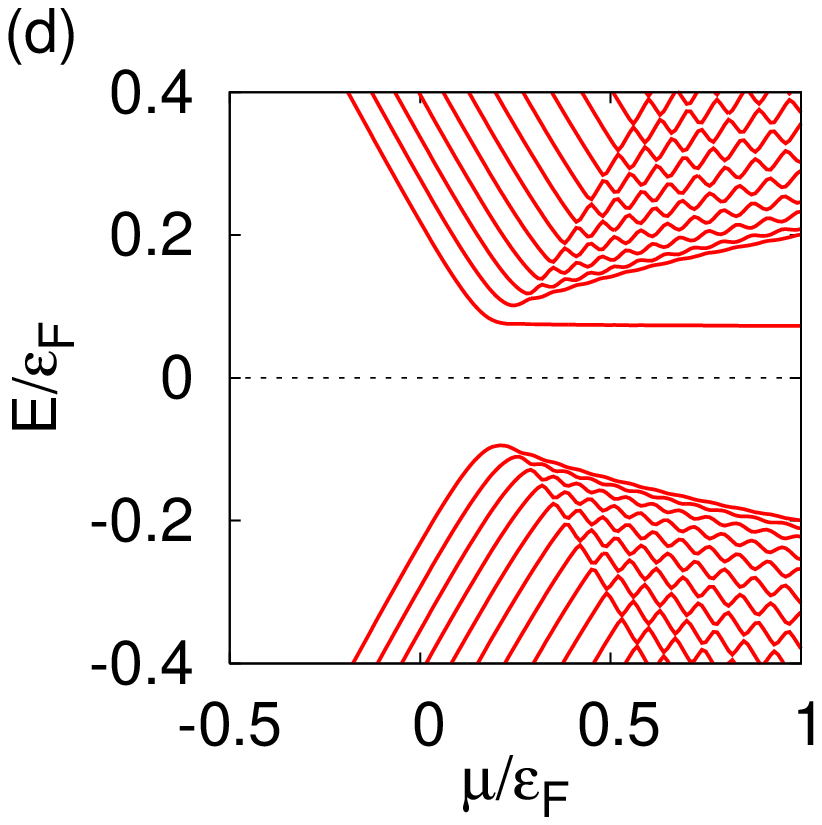}
\end{center}
\end{minipage}
\end{tabular}
\caption{Examples of spectrum in the $p+ip$-wave SF when
$k_FR=80,k_F\Delta=0.2\varepsilon_F$ for
(a) $\mu=-0.3\varepsilon_F$ (BEC regime),
(b) $\mu=0$, and (c) $\mu=\varepsilon_F$ (BCS regime). 
(d) Evolution of the spectrum
for fixed $l=-30$ as $\mu$ is changed.}
\label{fig:p+ip}
\end{figure}

We first discuss the $p+ip$-wave states ($\nu=1$) for which
the spectrum is particle-hole symmetric about $l=-1/2$:
\begin{align}
 \{ E^{(l)}_{m} \}_{m=1}^{2M} =
\{ - E^{(-l-1)}_m \}_{m=1}^{2M} .
\label{eq:PHS-Elm}
\end{align}
For simplicity,
we treat the two parameters $\mu$ and $\Delta$
independently as in Refs.~\cite{pap:Kita1998,pap:Stone2008,pap:Read2000}
for a discussion of the spectral flow.
In the BEC regime, by numerically diagonalizing $H_{\rm BdG}^{(l)}$,
we obtain fully gapped spectrum as shown in Fig. \ref{fig:p+ip}(a),
and find that the spectral asymmetry $\eta_l$
is zero for all $l$.
As a consequence, $\mathcal{L}_z=0$ and thus follows $\Lztot = N/2$,
as expected.
However, the spectrum becomes less trivial if the system is in the
BCS regime. 
There, a single edge mode with
$E_{\rm edge}^{(l)}\propto-(l+1/2)$ appears
as seen in Fig. \ref{fig:p+ip}(c),
refecting the topological nature
of the phase~\cite{pap:Read2000,book:Volovik2003,pap:Hasan2010,pap:Qi2011}.
This edge mode is particle-hole symmetric by itself,
$E_{\rm edge}^{(l)}\sim-(l+1/2) \sim - E_{\rm edge}^{(-l-1)}$,
and is hereafter
called a particle-hole symmetric (PHS) edge mode.
For spatially constant, non-vanishing pair potentials,
we have confirmed that $\eta_l = 0$ for all values of $l$,
for a sufficiently large system size where
the finite-size discretization of the energy levels
is small compared to 
$\Delta_0= k_F \Delta$ where $k_F$ is the Fermi wavenumber.
Namely, even deep inside the BCS regime, and even for
a small $\Delta_0$, 
we find $\mathcal{L}_z =0$ in the thermodynamic limit, which
implies no reduction of the OAM: $\Lztot = N/2$ holds exactly.

A natural question arising here is why $\mathcal{L}_z$ remains
zero in the BCS regime, despite the quantum phase transition
separating it from the BEC regime.
Let us first consider the limit $\mu=-\infty$ 
with a fixed $\Delta>0$, where $L_z=N=0$ holds trivially.
Thus we find $\calL_z=0$ in this limit.
Increasing $\mu$, $L_z$ and $N$ acquire non-zero values.
However, as long as there is no gap closing,
$\calL_z=0$ and thus $\Lztot = N/2$ still hold.
This gives a proof for the physical expectation that
$\Lztot = N/2$ holds throughout
the BEC regime, which belong to the non-topological
strong-pairing phase.

The value of $\calL_z$ in the BCS regime (weak-pairing phase)
is more subtle, due to the presence of
the quantum phase transition
at $\mu=0$ in the thermodynamic limit.
At the quantum phase transition, a gap closing is expected.
However, a careful examination reveals that
every eigenvalue keeps its sign when $\mu$ is continuously 
varied from $\mu=-\infty$ to $\mu=\varepsilon_F>0$.
In fact, for the system defined on a finite disc, the
gap never closes. The gap does approach zero
at the quantum critical point, and also inside the BCS regime
giving rise to the gapless chiral edge mode, but
only in the thermodynamic limit.
The change of the eigenvalues when $\mu$ is varied
is shown in Fig. \ref{fig:p+ip} (d). 
The edge mode appears in the BCS regime,
as a set of eigenstates separated from bulk states.
Although it converges to the linear gapless dispersion,
all the eigenvalues corresponding to the edge mode
for $l \leq -1$ come off from
the upper continuum of bulk eigenstates and remain positive.
Likewise, all the edge mode eigenvalues for $l\geq 0$
come from the lower continuum and remain negative.
Since all the eigenvalues depend continuously on $\mu$,
this is the only possible evolution to generate
the PHS edge mode, under
the particle-hole symmetry~\eqref{eq:PHS-Elm}.

Although each of $L_z$ and $N$ changes from the trivial values
$L_z=N=0$, ${\mathcal L}_z=L_z-N/2=0$ is always
satisfied and any correction factor like $(\Delta_0/\varepsilon_F)^{\gamma}$
discussed in the introduction cannot arise.
We note that this is also true for a physical process where $\mu$ and
$\Delta$ are simultaneously tuned to keep $N$ constant.
Our argument only relies on the formation of
the PHS edge mode separated from the bulk eigenstates,
which is valid for a sufficiently large system size.
The conclusion of our analysis largely agrees with the recent related
calculations on
$p+ip$-wave SF~\cite{pap:Stone2008,pap:Sauls2011,pap:Mizushima2008,
pap:Tsutsumi2012PRB},
but clarify why $\Lztot$ is exactly given by 
$N/2$ even for small $\Delta_0/\varepsilon_F$.

\begin{figure}[htbp]
\begin{tabular}{ccc}
\begin{minipage}{0.33\hsize}
\begin{center}
\includegraphics[width=\hsize,height=\hsize]{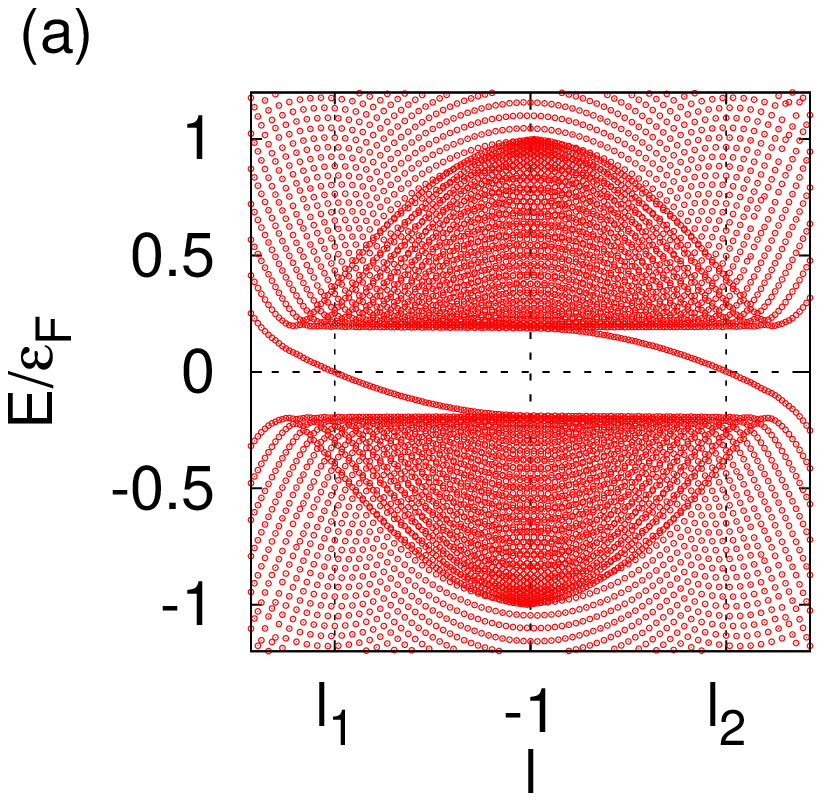}
\end{center}
\end{minipage}
\begin{minipage}{0.33\hsize}
\begin{center}
\includegraphics[width=\hsize,height=\hsize]{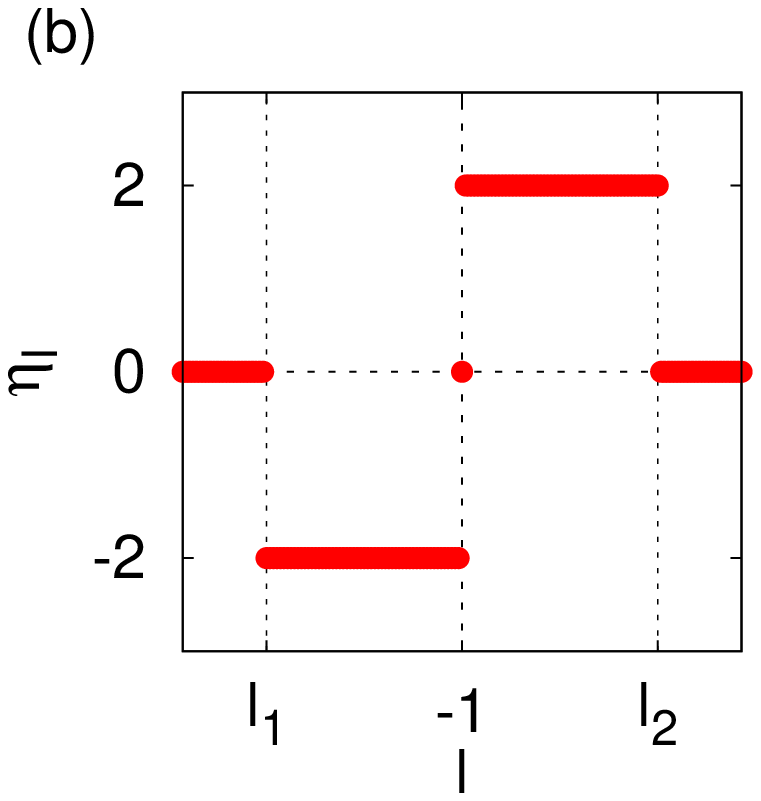}
\end{center}
\end{minipage}
\begin{minipage}{0.33\hsize}
\begin{center}
\includegraphics[width=\hsize,height=\hsize]{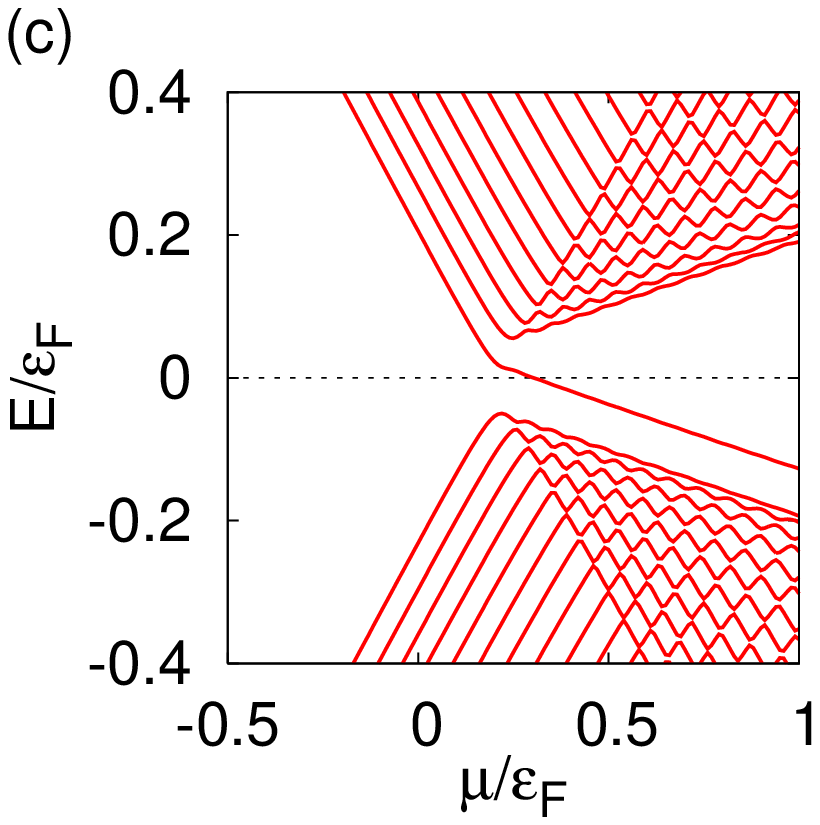}
\end{center}
\end{minipage}
\end{tabular}
\caption{Examples of (a) spectrum in the $d+id$-wave SF 
and (b) the spectral asymmetry $\eta_l$ for
$k_FR=80,k_F^2\Delta=0.2\varepsilon_F,
\mu=\varepsilon_F$ (BCS regime).
(c) Spectral flow
for fixed $l=-30(>l_1=-56)$ as $\mu$ is changed.}
\label{fig:d+id}
\end{figure}

Next, we move to the $d+id$-wave states for which 
the spectrum is particle-hole symmetric about $l=-1$.
We have numerically confirmed that $\eta_l=0$ for all $l$
in the BEC regime and obtain ${\mathcal L}_z=0$, i.e. $L_z=N$.
On the other hand, in the BCS states,
it is known that a $d+id$-wave state
has two non-degenerate edge modes at one boundary as visualized in Fig. 
\ref{fig:d+id} (a)
\cite{pap:Yang1994}.
Each edge mode 
is particle-hole symmetric with the other branch,
but not symmetric by itself; we call them non-PHS edge modes.
Their dispersion relations are given as
$E_{{\rm edge}1,2}^{(l)}\propto-(l-l_{1,2})$,
where $l_{1,2} \neq -1(l_1<l_2)$ are
the ``Fermi angular momenta'' where the edge modes cross
the zero energy.
The PHS of the system requires $l_1+1=-(l_2+1)$.
Interestingly, for this spectrum, 
we find that some of the $\eta_l$'s become non-vanishing,
as shown in Fig. \ref{fig:d+id}(b):
$\eta_l=0$ for $l<l_1$, $l>l_2$, and $l=-1$,
while $\eta_l=-2$ for $l_1<l<-1$ and $\eta_l=+2$ for $-1<l<l_2$.

This result can be understood in terms of the spectral flow
staring from the BEC regime where $\eta_l=0$.
As in the case of the $p+ip$-wave SF, $\eta_l=0$ remains valid
up to the critical point, since the gap remains open in the
entire BEC regime.
At the critical point, the gap in the spectrum is minimal at $l=-1$
as shown in Fig.~\ref{fig:p+ip} (b), corresponding to the closing
of the gap.
However, the gap in a finite-size system does not vanish at $l=-1$.
As we move into the BCS regime,
two non-PHS edge modes develop as in
Fig.~\ref{fig:d+id} (a).
During this evolution, the Fermi angular momenta
evolves from $l=-1$ to non-vanishing values $l_{1,2}$.
This induces spectral flow for the angular momenta $l$ in the
range $l_1 < l < l_2$, except at $l=-1$.
An example of the spectral flow at a fixed value of $l$ is
shown in Fig. \ref{fig:d+id} (c): $\eta_l$ changes sign
exactly when a Fermi angular momentum passes through
this $l$.
This picture can be also confirmed in an analytic expression of
the edge mode dispersions as functions of $\mu>0$, 
as follows~\cite{com:supple}. 
Let us consider the limit of
a large disc radius $R$.
The edge of the disc then corresponds to the boundary of
a semi-infinite plane, where the momentum parallel to
the boundary, $k_{\parallel}$, is related to
the angular momentum $l$ by
$k_{\parallel} \simeq l/R.$
In the BCS regime $0 < \mu$,
the dispersion of the edge modes are given as
$E^2_{{\rm edge}}=(\Delta_0^2\varepsilon_F^2/(\varepsilon_F^2+\Delta_0^2))
(2k_{\parallel}^2/k_F^2-\mu/\varepsilon_F)^2$ where $\Delta_0=k_F^2\Delta$
by solving the BdG equation~\cite{com:supple}.
The condition $E^2_{\mbox{\scriptsize edge}}=0$ then
determines the Fermi wavevector $k_{F\parallel}$
of the edge modes as
\begin{align}
k_{F\parallel}^2=
\left(\frac{k_F}{\sqrt{2}}\right)^2\frac{\mu}{\varepsilon_F}.
\end{align}
This indeed demonstrates that the Fermi wavevector
$k_{F\parallel}$ evolves from
zero to a non-vanishing value, as $\mu$ is increased from
the critical point $\mu=0$ into the BCS regime.
This confirms the spectral flow for $l_1<l<l_2$ 
as found numerically.
In fact, spectral flows, and $\eta_l\neq 0$ as a consequence,
are common properties of the 
higher-order pairing states with $\nu\geq2$ which have non-PHS edge modes
in the BCS regime. 
We have also numerically calculated $\eta_l$ for $\nu=3,4$.
As expected,
we find $\eta_l=0$ for all $l$ in the BEC regime.
In the BCS regime, on the other hand, there appear $\nu$ edge
modes: one PHS edge mode and $\nu-1$ non-PHS edge modes for an odd $\nu$,
and $\nu$ non-PHS edge modes for an even $\nu$.
Let $\{l_j\} (l_j<l_{j+1})$
be Fermi angular momenta for these edge modes, namely
the edge modes cross zero energy at $l=l_j$.
For $\nu=3$, we find $\eta_l=\pm2$ if 
$|l| <|l_{1,3}|$
and $\eta_l=0$ otherwise;
for $\nu=4$,
$\eta_l=\pm 2$ if $l_{1,3}<l<l_{2,4}$, $\eta_l=\pm 4$ if
$l_2<l< -2$ or $-2 < l < l_3$, and $\eta_l=0$ otherwise.
These results can be naturally understood in terms of
the spectral flow starting from $\mu = - \infty$ deep in
the BEC regime, as in the case of $\nu=2$ discussed above.
This pattern is expected to persist for any $\nu \geq 2$.
To summarize, 
in general chiral SF with $\nu \geq 2$, $\eta_l$ changes by $\pm2$ when
a non-PHS edge mode branch crosses zero energy,
while a PHS edge mode does not contribute to $\eta_l$.

In the BCS regime for $\nu\geq2$,
non-vanishing $\eta_l$ implies
${\mathcal L}_z \neq 0$.
As a consequence, $L_z$ is in fact strongly suppressed
from the ``full'' value $\nu N/2$, in the BCS limit 
$\Delta_0= k_F^{\nu}\Delta\ll \varepsilon_F$.
To see this, we evaluate the actual value of $\calLz$
using Eq.~\eqref{eq:L}, with the observations made above.
By considering the limit of a
large disc, 
the ``Fermi angular momenta'' $l_j$ can be written in terms of
the Fermi wavenumber parallel to the boundary
$k_{F\parallel}^{(j)}$ as $l_j \simeq R k_{F\parallel}^{(j)}$.
Within the quasi-classical formulation, which is legitimate for
the BCS limit,
we find
$\sum_{j=1}^{\nu}(k_{F\parallel}^{(j)})^2 
= \nu k_F^2/2$
~\cite{com:supple}.
Thus, in the leading order in $N$ and $\Delta_0/\varepsilon_F$, 
we obtain
\begin{align}
 \calLz \simeq -\frac{1}{2} \sum_{j=1}^\nu {l_j}^2 =
- \frac{1}{2} \sum_{j=1}^\nu \left(R k_{F \parallel}^{(j)}\right)^2
=- \frac{\nu N}{2}. 
\end{align}
Since ${\mathcal L}_z=L_z-\nu N/2$, the OAM
is evaluated to be
$L_z=N\times O(\Delta_0/\varepsilon_F)$ 
in the BCS limit with $\nu\geq 2$,
where the  $O(\Delta_0/\varepsilon_F)$ term represents
possible additional contributions which are beyond
the quasi-classical approximation.
Indeed, numerical calculations of the OAM give
$L_z/N\sim o(0.01)$ when $N\sim O(1000)$
for $\nu=2,3,4$ in an extended range of
the BCS regimes with $\Delta_0/\varepsilon_F
\lesssim0.2$, supporting the above quasi-classical analysis.
Therefore, 
the naive evaluation $L_z=\nu N/2$ fails for the chiral SFs
in the BCS regime with $\nu\geq2$, 
even though it gives the correct value for the $p+ip$-wave states.
That is, $L_z$ is strongly suppressed as if
in the naive weak-pairing picture where fermions only near the
Fermi surface at $\Delta=0$ contribute to $L_z$,
but only for $\nu \geq 2$.
However,
our findings, in particular the stark difference between
the $p+ip$ and higher-order ($\nu \geq 2$) pairing cases,
make it clear that
the suppression cannot be understood by any of the arguments
found in existing works.
Our analysis is based on the
robustness of the spectral asymmetry $\eta_l$,
and does not rely on assumptions and 
approximations used in the eariler papers,
such as derivative expansions, 
which might fail to describe the correct physics especially around
boundaries where the edge modes are
located~\cite{pap:Sauls2011}. 
We emphasize that
the well-known topological protection of existence of $\nu$ edge modes
is not sufficient for determining the
OAM, which depends on more detailed structures of the edge modes.
In this sense, the OAM $L_z$ 
is a surface-dependent quantity
in the BCS regimes.

Finally, let us discuss why the OAM is suppressed for $\nu \geq 2$
but not for $\nu=1$, in terms of the ground-state wave function.
A general expression~\cite{com:supple,pap:Lanbote1974}
for the ground state of a BdG Hamiltonian is given as
$|{\rm GS}\rangle ={\mathcal N}\otimes_l |{\rm GS}
\rangle_l$, where
\begin{align}
|{\rm GS}\rangle_l&=\left(\prod_{j=1}^{n_{\uparrow}^{(l)}}
\tilde{c}_{j,l+\nu,\uparrow}^{\dagger}\right)
\left(\prod_{j=1}^{n_{\downarrow}^{(l)}}
\tilde{c}_{j,-l,\downarrow}^{\dagger}\right)\notag\\
&\times 
\exp \left(\sum_{j>n_{\uparrow}^{(l)}}\sum_{j^{\prime}>n_{\downarrow}^{(l)}}
\tilde{c}_{j,l+\nu,\uparrow}^{\dagger}
F^{(l)}_{jj^{\prime}}
\tilde{c}_{j^{\prime},-l,\downarrow}^{\dagger}\right)|0\rangle.
\label{eq:GS}
\end{align}
Here $|0\rangle$ is the vacuum for $c_{nl\sigma}$ and ${\mathcal N}$ is a
normalization constant,
$\tilde{c}_{jl\sigma}$ is a linear superposition of 
$\{c_{nl\sigma}\}_n$, 
and $n_{\uparrow}^{(l)}$, $n_{\downarrow}^{(l)}$ are non-negative integers.
The ground state of a BdG Hamiltonian is often assumed to have
a pure exponential form ($n_\sigma^{(l)}=0$ in Eq.~\eqref{eq:GS}),
which implies that all the
fermions are paired and thus $\Lztot = \nu N/2$.
For $\nu=1$, the ground state (of a sufficiently large system)
is indeed reduced to the pure exponential form, implying the
full OAM $\Lztot = N/2$.
However, the ground state of a BdG Hamiltonian generally takes
the form of Eq.~\eqref{eq:GS}.
A non-vanishing $n_\sigma^{(l)}$ signals the existence of
\emph{unpaired} fermions, which contribute to the reduction of the OAM.

In fact, we can derive~\cite{com:supple} the identity
\begin{align}
{\mathcal L}_z&=-\frac{1}{2}\sum_l\left(l+\frac{\nu}{2}\right) n^{(l)},
& n^{(l)}&=2\left(n_{\downarrow}^{(l)}
-n_{\uparrow}^{(l)}\right),
\label{eq:L2}
\end{align}
which explicitly shows that the unpaired fermions are necessary
for the reduction of the OAM, 
which requires $\mathcal{L}_z \neq 0$.
Furthermore, the reduction requires a non-vanishing difference
$n_\uparrow^{(l)} - n_\downarrow^{(l)}$.
Indeed, when $n_{\uparrow}^{(l)}= n_{\downarrow}^{(l)}$,
$|{\rm GS}\rangle$ belongs to the eigenvalue ${\mathcal L}_z=0$.
This is because the vacuum $|0\rangle$ belongs to $\calLz=0$
and every creation operator in Eq.~\eqref{eq:GS}
comes in the pair
$\tilde{c}_{l+\nu\uparrow}^{\dagger}
\tilde{c}_{-l\downarrow}^{\dagger}$,
which commutes with $\hat{\mathcal L}_z$.

Comparing Eqs.~\eqref{eq:L} and~\eqref{eq:L2},
we obtain $n^{(l)}=\eta_l$.
Together with the property $n_\uparrow^{(l)} n_\downarrow^{(l)} =0$,
we find the numbers of unpaired fermions explicitly as follows
~\cite{com:supple}:
For $\nu=1$, $n_\uparrow^{(l)} = n_\downarrow^{(l)} =0$, namely
there is no unpaired fermions as mentioned earlier.
For $\nu \geq 2$, $n_{\uparrow}^{(l)}>0, n_{\downarrow}^{(l)}=0$
for $l_1<l<-\nu/2$, and $n_{\uparrow}^{(l)}=0, n_{\downarrow}^{(l)}>0$
for $-\nu/2<l<l_{\nu}$.
Therefore, the unpaired fermions generally carry angular momentum 
{\it opposite} to the given chirality, and this contribution
cancels out with that from the paired fermions,
leading to the reduction of $L_z$ from the full value $\nu N/2$.
This depairing effect is
associated with the formation of the 
non-PHS edge modes, signifying the fundamental difference
between $\nu=1$ and $\nu \geq 2$ cases.

We thank Y. Maeno for introducing the problem of
the OAM in chiral superfluids to us.
We are also grateful to 
E. Demler, S. Fujimoto, A. Furusaki, J. Goryo, 
O. Ishikawa, N. Kawakami, Y. B. Kim, 
T. Kita, T. Mizushima, S.-J. Mao, E.-G. Moon, Y. Nishida,
M. Sato, K. Shiozaki, A. Shitade, M. Sigrist, H. Sumiyoshi, 
W.-F. Tsai, Y. Tsutsumi, K. Ueda, and S.-K. Yip
for valuable discussions.
This work was supported by the ``Topological Quantum Phenomena''
(No. 25103706) Grant-in Aid for Scientific Research on Innovative Areas
from the Ministry of Education, Culture, Sports, Science and Technology
(MEXT) of Japan.




\renewcommand{\bibnumfmt}[1]{[S#1]}
\renewcommand{\theequation}{S\arabic{equation}}
\renewcommand{\thefigure}{S\arabic{figure}}
\renewcommand{\thetable}{S.\Roman{table}}

\setcounter{equation}{0}
\setcounter{figure}{0}
\setcounter{table}{0}

\clearpage
\onecolumngrid
\section{Supplemental Material}
\twocolumngrid


\subsection{A. Analytic Solution for Edge Modes}
Analytic solutions for the edge mode branches on a semi-infinite plane
can be obtained by a straightforward calculation.
We consider a semi-infinite plane where $y<0$ is a superfluid region
and $y>0$ is vacuum.
Our Hamiltonian for $y<0$ reads
\begin{align}
H(k_x)=\left[
\begin{array}{cc}
(k_x^2-\partial_y^2)/2m_0-\mu & \Delta(k_x+\partial_y)^{\nu}\\
\Delta(k_x-\partial_y)^{\nu} & -(k_x^2-\partial_y^2)/2m_0+\mu 
\end{array}
\right].
\end{align}
Analytic properties of a similar Hamiltonian with $\nu=1$ have been discussed 
in the context of the topological insulators
~\cite{pap:Zhou2008s,pap:Wada2011s}.
We first discuss the $p+ip$-wave SF and then move to the
$d+id$-wave SF.
The edge mode wavefunction $u$ with 
a boundary condition $u(k_x,y=0)=0$
is given by
\begin{align}
u&=\left[
\begin{array}{c}
a\\b
\end{array}
\right](e^{\lambda_1y}-e^{\lambda_2y}),\\
\lambda_{1,2}^2&=k_x^2+2m_0\left(
\frac{2m_0\Delta^2}{2}-\mu\right)\notag \\
&\quad
\pm2m_0\sqrt{\left(\frac{2m_0\Delta^2}{2}-\mu\right)^2
-\mu^2+E^2},
\label{eq:lam}
\end{align}
where Re$\lambda_{1,2}>0$ so that $u(k_x,y\rightarrow-\infty)=0$.
Since $Hu=Eu$, non-zero $a,b$ lead to two equations
$(\varepsilon_1-E)(k_x+\lambda_2)=(\varepsilon_2-E)(k_x+\lambda_1)$ and
$(\varepsilon_1+E)(k_x-\lambda_2)=
(\varepsilon_2+E)(k_x-\lambda_1)$,
where $\varepsilon_{1,2}=(k_x^2-\lambda_{1,2}^2)/2m_0-\mu.$
By solving these equations, we obtain
\begin{align}
\lambda_{1}+\lambda_2&=2m_0E/k_x,\\
\lambda_{1}\lambda_2&=2m_0\mu-k_x^2.
\label{eq:lam_}
\end{align}
From Eqs. (\ref{eq:lam}) and (\ref{eq:lam_}),
the edge mode energy is given as
\begin{align}
[E(k_x)]^2=\Delta_0^2k_x^2/k_F^2.
\end{align}
Note that this is valid only for Re$\lambda_{1,2}>0$,
or equivalently $k_x^2<k_F^2\mu/\varepsilon_F$.
From $\lambda_{1,2}$ and $E$ together with the normalization of $u$, 
the components of $u$
are determined as $a^2=b^2=\Delta_0/v_F=1/\xi$ where $v_F=k_F/m_0$.

For the $d+id$-wave SF, we can compute the edge mode branches
in the same way.
In this case, $\lambda_{1,2}$ and $E$ are determined by
\begin{align}
&\lambda_{1,2}^2=-\frac{\mu/(2m_0)}{[1/(2m_0)]^2+\Delta^2}+k_x^2\notag\\
&\qquad \pm\frac{1}{\sqrt{[1/(2m_0)]^2+\Delta^2}}
\sqrt{\frac{-\mu^2\Delta^2}{[1/(2m_0)]^2+\Delta^2}+E^2},\\
&\lambda_{1}+\lambda_2=\frac{2k_xE}{2k_x^2/(2m_0)-\mu},\\
&\lambda_{1}\lambda_2=2m_0\mu+\frac{2m_0E^2}{2k_x^2/(2m_0)-\mu}-k_x^2.
\end{align}
From these equations, we obtain
\begin{align}
[E_{1,2}(k_x)]^2=\frac{\Delta_0^2\varepsilon_F^2}{\varepsilon_F^2+\Delta_0^2}
\left(2\frac{k_x^2}{k_F^2}-\frac{\mu}{\varepsilon_F}\right)^2.
\label{eq:Ed}
\end{align}
The above $E$ is reduced to the well-known quasi-classical result
$E_{1,2}^2=\Delta_0^2(2k_x^2/k_F^2-1)^2$ when $\mu=\varepsilon_F\gg\Delta_0$
~\cite{pap:Yang1994s}.
The components $a_{1,2},b_{1,2}$ are the solutions of
\begin{align}
a_j^2+b_j^2&=\frac{2\Delta_0|k_x|}{\sqrt{\varepsilon_F^2+\Delta_0^2}}
\frac{\mu\varepsilon_F-(\varepsilon_F^2-\Delta_0^2)k^2/k_F^2}
{\mu\varepsilon_F-(\varepsilon_F^2+\Delta_0^2)k^2/k_F^2},\label{eq:a2b2}\\
\frac{a_1^2}{b_1^2}&=\frac{\varepsilon_F^2+\Delta_0^2}{\varepsilon_F^2}
\left(1+\frac{\Delta_0}{\sqrt{\varepsilon_F^2+\Delta_0^2}}\right)^2,\\
\frac{a_2^2}{b_2^2}&=\frac{\varepsilon_F^2+\Delta_0^2}{\varepsilon_F^2}
\left(1-\frac{\Delta_0}{\sqrt{\varepsilon_F^2+\Delta_0^2}}\right)^2.
\end{align}
It is easy to check $a_1^2=b_2^2$ and $b_1^2=a_2^2$ which are 
implied from the particle-hole symmetry of the Hamiltonian,
$PH(k_x)P^{-1}=-H(-k_x)$ with $P=i\sigma_y$.
Note that $k_x$ is restricted to 
$0<k_x^2<k_F^2\mu\varepsilon_F/(\varepsilon_F^2+\Delta_0^2)$
from the condition Re$\lambda_{1,2}>0$ and Eq. (\ref{eq:a2b2}).
In the quasi-classical limit, $a_j,b_j$ are simply reduced to
$a_j^2=b_j^2=2|k_x|/\xi k_F$.

\subsection{B. Quasi-Classical Calculation of ${\mathcal L}_z$}
Here, we discuss ${\mathcal L}_z$ in 
the quasi-classi limit 
$\Delta_0=k_F^{\nu}\Delta\ll \varepsilon_F$ in the BCS regimes.
First, as we discussed in the main text and saw in Fig. 2 (b), 
$\eta_l$ changes $\pm$2 only when a non-PHS 
edge mode branch crosses zero energy
as a function of $l$, which is confirmed numerically.
Therefore, we simply count how many
times the non-PHS edge mode branches cut zero energy, and obtain
\begin{align}
\eta_l=-2\sum_{j=1}^{[\nu/2]}\theta(l-l_j)
\end{align}
for $l< -\nu/2$,
where $l_j$ is the ``Fermi angular momenta'' at which 
$j$-th edge mode branch crosses zero energy as a function of
$l$ and
$\theta(l-l_j)$ is the step function.
We have labeled the edge modes so that $l_1<l_2<\cdots <l_{\nu}$
is satisfied.
$\eta_{l}=-\eta_{-l-\nu}$ holds for 
$l>-\nu/2$ by the particle-hole symmetry.
By using this, we obtain
\begin{align}
{\mathcal L}_z&=-\frac{1}{2}\times 2\sum_{j=1}^{[\nu/2]}
2\times \frac{1}{2}(l_j+\nu/2)(l_j+\nu/2-1)\notag \\
&\simeq -\frac{1}{2}\sum_{j=1}^{\nu}l_j^2
\label{eq:Ll}
\end{align}
for even $\nu$s.
In the second equality, we neglected $o(l_j)$ contributions
which is much smaller than the $l_j^2$-contributions
when the total number of fermions $N$ is large enough.
Contributions from $l>-\nu/2$ has been included by introducing
a factor 2 in front of the summation $\sum_{j=1}^{[\nu/2]}$.
Equation (\ref{eq:Ll}) also holds for odd $\nu$s.

The Fermi angular momentum $\{l_j\}$ of the edge modes
are evaluated within quasi-classical calculations.
When the radius of the system is large,
$\xi=v_F/\Delta_0\ll R\rightarrow \infty$,
curvature of the boundary could be negligible and the physics 
around the boundary would be equivalent to that in a
cylinder system or a semi-infinite system.
In such systems, 
edge mode wavefunctions are proportional to 
$\exp[ik_{\parallel}^{}x_{\parallel}]$ where 
$k_{\parallel}^{},x_{\parallel}$
are a wavenumber and a position along the boundary, respectively.
Since the edge modes are well localized around the boundary,
we can regard the edge modes as running at the boundary $r=R$
in the disc geometry. 
In this case, the (anti-)periodic boundary condition is satisfied,
$k_{\parallel}^{}\simeq 2\pi l/(2\pi R)=l/R$.
Especially, the Fermi angular momentum $\{l_j\} $ of the edge modes
are connected with their Fermi wavenumbers $\{k_{F\parallel}^{(j)}\}$ as
\begin{align}
l_j\simeq Rk_{F\parallel}^{(j)}.
\end{align}
Therefore, we obtain a quasi-classical expression for ${\mathcal L}_z$,
\begin{align}
{\mathcal L}_z\simeq-\frac{1}{2}\sum_{j=1}^{\nu}(Rk_{F\parallel}^{(j)})^2.
\label{eq:quasi_L}
\end{align}

The Fermi wavenumbers $\{k_{F\parallel}^{(j)}\}$ can easily be
calculated for semi-infinite systems within the quasi-classical calculations.
Let us assume that the system has a straight boundary along $y$-axis
at $x=0$ which separates the SF in $x<0$ from the vacuum in $x>0$.
By solving the BdG equation near the boundary,
we obtain an equation which determines the edge mode energy $E$
\cite{pap:Yang1994s},
\begin{align}
\frac{\Delta_{+}}{E-i\sqrt{|\Delta_+|^2-E^2}}
=\frac{\Delta_{-}}{E+i\sqrt{|\Delta_-|^2-E^2}}.
\end{align}
Here, we have defined $\Delta_{\pm}=\Delta_0 (\pm k_x+ik_y)^{\nu}
/k_F^{\nu}$ with $k_x=\sqrt{k_F^2-k_y^2}$.
If we rewrite $\Delta_+=\Delta_{\rm odd}+\Delta_{\rm even}$ where
$\Delta_{\rm odd/even}$ are odd/even with respect to $k_x\leftrightarrow -k_x$,
the above equation leads to
$\Delta_{\rm even}=0$ at the Fermi wavenumber of the one dimensional edge modes
where $E=0$.
By using $k_x=k_F\cos \theta, k_y=k_F\sin\theta$,
it is seen that $\Delta_{\rm even}=
\Delta_0\cos (\nu\theta)$ for an even $\nu$, and
$\Delta_{\rm even}=i
\Delta_0\sin (\nu\theta)$ for an odd $\nu$.

Then, we can evaluate the Fermi wavenumbers of the $\nu$ 
edge modes along the $y$-direction.
It is written as $k_{F\parallel}^{(j)}=k_F\sin \theta_F^{(j)}$ where 
$\theta_F^{(j)}=-\pi/2+(2j-1)\pi/2\nu$ for $j=1,\cdots,\nu$
satisfying 
$\Delta_{\rm even}(\theta_F^{(j)})=0$.
For $\nu\geq2$, since
$\sum_{j=1}^{\nu}e^{i(2j-1)\pi/\nu}=0,$
we obtain an identity within the quasi-classical
calculation,
\begin{align}
\sum_{j=1}^{\nu}\left(k_{F\parallel}^{(j)}\right)^2
&=\sum_{j=1}^{\nu}k_F^2[\sin(\theta_{F}^{(j)})]^2\notag\\
&=k_F^2\sum_{j=1}^{\nu}\left(\frac{1}{2}-\frac{\cos(2\theta_{F}^{(j)})}
{2}\right)\notag\\
&=\nu \frac{k_F^2}{2}.
\label{eq:quasi_id}
\end{align}
There would be possible deviations of $k_{F\parallel}^{(j)}$ 
from the above values,
which are beyond the present quasi-classical approximation.
From Eq. \eqref{eq:quasi_id}, we obtain the same identity
for the perpendicular component 
$k_{F\perp}^{(j)}=k_F\cos \theta_F^{(j)}$,
\begin{align}
\sum_{j=1}^{\nu}\left(k_{F\perp}^{(j)}\right)^2
&=\nu \frac{k_F^2}{2}.
\label{eq:quasi_id2}
\end{align}
These equations are a direct consequence of the residual $\nu$-hold
rotational symmetry of the gap function.
The condition $\Delta_{\rm even}=0$ means vanishing edge mode energy $E=0$
when the two-dimensional wavenumber $(k_{F\perp}^{(j)},k_{F\parallel}^{(j)})$
is at nodes of $\Delta_{\rm even}$, where positions of the nodes
are $\nu$-hold rotationally symmetric.
This symmetry immediately leads to $\sum_{j=1}^{\nu}(k_{F\perp}^{(j)})^2
=\sum_{j=1}^{\nu}(k_{F\parallel}^{(j)})^2$, from which
we can easily reproduce Eqs. \eqref{eq:quasi_id} and \eqref{eq:quasi_id2}
since the wavenumber is on the two-dimensional Fermi surface,
$[(k_{F\perp}^{(j)})^2+(k_{F\parallel}^{(j)})^2]=k_F^2$ for each $j$
within the quasi-classical calculation.
On the other hand, for $\nu=1$, there is no residual rotational symmetry
in $\Delta_{\rm even}$.
This explains the difference between $\nu=1$ and $\nu\geq2$.

We can see that Eqs. \eqref{eq:quasi_id} and \eqref{eq:quasi_id2} 
are indeed satisfied by directly looking at 
$k_{F\parallel}^{(j)}$; 
$|k_{F\parallel}^{(j)}|=k_F/\sqrt{2}\simeq
0.71k_F$ for $\nu=2$,
$|k_{F\parallel}^{(j)}|=0, \sqrt{3}k_F/2\simeq 0.87k_F$ for $\nu=3$, and
$|k_{F\parallel}^{(j)}|=k_F\sqrt{1/2\pm1/2\sqrt{2}}\simeq 
0.93k_F,0.38k_F$ for $\nu=4$, and so on.
We have numcerically confirmed that
Eq. \eqref{eq:quasi_id} holds also for 
the disc geometry. For example, when $k_FR=80$,
the edge mode Fermi wavenumbers are numerically obtained as
$|k_{F\parallel}^{(1,2)}|\simeq56k_F/80\simeq 0.70k_F$ 
for $\nu=2$,
$|k_{F\parallel}^{(1,3)}|\simeq68k_F/80\simeq 0.85k_F,
|k_{F\parallel}^{(2)}|\simeq1k_F/80\simeq 0.013k_F$ for $\nu=3$,
and 
$|k_{F\parallel}^{(1,4)}|\simeq72k_F/80\simeq 0.90k_F,
|k_{F\parallel}^{(2,3)}|\simeq31k_F/80\simeq 0.39k_F$ for $\nu=4$
for several values of $\Delta_0\ll \varepsilon_F$.
We note that,
although the above discussions are not based on self-consistent calculations,
Eq. \eqref{eq:quasi_id} should be satisfied even in self-consistent 
calculations for the BCS limit,
since $\{k_{F\parallel}^{(j)}\}$ are generally determined by symmetry of the
gap functions.

When $\mu$ and $\Delta$ are tuned so that 
$N$ is fixed,
the total number of fermions on the disc is expressed as
\begin{align}
N&\simeq Vn_{2D}
=R^2k_F^2/2, 
\label{eq:BCS_N}
\end{align}
where $n_{2D}=k_F^2/2\pi$ is the particle density in
two dimensions and $V=\pi R^2$ is the volume of the system.
We have neglected $o(R)$ contributions which can
arise from deviations of the particle density from $n_{2D}$
near the boundary.
From Eqs. (\ref{eq:quasi_L}), (\ref{eq:quasi_id}), and
(\ref{eq:BCS_N}), 
we obtain Eq. (6) in the main text.

\subsection{C. Ground State Wavefunctions}
A general theory of the Bogoliubov transformation was developed 
in Ref. \cite{pap:Lanbote1974s}.
We follow it and obtain 
the ground state wavefunctions for the chiral SFs on the disc.
The Bogoliubov transformation is introduced as
\begin{align}
\left[
\begin{array}{c}
c_{nl+\nu\uparrow}\\
c_{n-l\downarrow}^{\dagger}\\
\end{array}
\right]
=\sum_{m=1}^{M}\left[
\begin{array}{cc}
u_{nm}^{(l)} & u_{nM+m}^{(l)}\\
v_{nm}^{(l)} & v_{nM+m}^{(l)}\\
\end{array}
\right]
\left[
\begin{array}{c}
b_{m}^{(l)}\\
b_{M+m}^{(l)}\\
\end{array}
\right],
\end{align}
where $(u,v)^T$ are eigenvectors of the BdG
equation,
\begin{align}
\sum_{n'=1}^{M}
\left(H_{{\rm BdG}}^{(l)}\right)_{nn'}
\left[
\begin{array}{c}
u_{n'm}^{(l)}\\
v_{n'm}^{(l)}\\
\end{array}
\right]=
E_{m}^{(l)}
\left[
\begin{array}{c}
u_{nm}^{(l)}\\
v_{nm}^{(l)}\\
\end{array}
\right].
\end{align}
We have introduced the cut-off $M\gg 1$
as in the main text. 
The ground state $|{\rm GS}\rangle$ 
is defined as a vacuum for all the quasi-particle
excitations with positive energies,
\begin{align}
\left\{
\begin{array}{ll}
b_m^{(l)}|{\rm GS}\rangle =0 & (E_{m}^{(l)}>0),\\
b_m^{(l)\dagger}|{\rm GS}\rangle =0 & (E_{m}^{(l)}<0).
\end{array}
\right.
\end{align}

Let the number of the positive (negative) eigenvalues 
of $H_{\rm BdG}^{(l)}$ be $n_{+}^{(l)}(n_-^{(l)})$ which
are $n_{+}^{(l)}\neq n_{-}^{(l)}$ in general,
and we label the eigenvalues so that $E^{(l)}_{1}\geq \cdots 
\geq E^{(l)}_{2M}$.
Correspondingly, the $b$-operators are rewritten as
$b_{m+}^{(l)}\equiv b^{(l)}_{m}$ for $E^{(l)}_{m}>0$
and $b_{m-}^{(l)}\equiv b^{(l)\dagger}_{m}$ for $E^{(l)}_{m}<0$.
We also rewrite the unitary matrix for the transformation as,
\begin{align}
U^{(l)}\equiv\left[
\begin{array}{cc}
u_{nm}^{(l)} & u_{nM+m}^{(l)}\\
v_{nm}^{(l)} & v_{nM+m}^{(l)}\\
\end{array}
\right]^{-1}
=\left[
\begin{array}{cc}
U^{(l)}_1 & U^{(l)}_2\\
U^{(l)}_3 & U^{(l)}_4\\
\end{array}
\right],
\end{align}
where sizes of the matrices are $n_{+}^{(l)}\times M$
for $U^{(l)}_{1,2}$ and
$n_{-}^{(l)}\times M$ for $U^{(l)}_{3,4}$.

Since $U^{(l)\dagger}_{3}U^{(l)}_3$ is an $M\times M$ Hermitian matrix,
it has real eigenvalues $\{\lambda^{(l)}_{3j}\}_{j=1}^{M}$
with eigenvectors $\{X^{(l)}_{3j}\}_{j=1}^M$,
\begin{align}
\left(U^{(l)\dagger}_{3}U^{(l)}_3\right)
X^{(l)}_{3j}=\lambda^{(l)}_{3j}X^{(l)}_{3j}.
\end{align}
We denote the number of $\lambda^{(l)}_{3j}=1$ as $n^{(l)}_{\uparrow}\geq0$.
$U^{(l)}_{3}U^{(l)\dagger}_3$ is a $n^{(l)}_-\times n^{(l)}_-$ Hermitian
matrix and has
real eigenvalues $\{\lambda^{(l)}_{1j}\}_{j=1}^{n^{(l)}_-}$
with eigenvectors $\{X^{(l)}_{1j}\}_{j=1}^{n^{(l)}_-}$.
Similarly, 
an $M\times M$ Hermitian matrix $U^{(l)\dagger}_{2}U^{(l)}_2$
has 
real eigenvalues $\{\lambda^{(l)}_{2j}\}_{j=1}^{M}$
with eigenvectors $\{X^{(l)}_{2j}\}_{j=1}^{M}$,
and a $n^{(l)}_{+}\times n^{(l)}_{+}$ Hermitian matrix 
$U^{(l)}_{2}U^{(l)\dagger}_2$
has 
real eigenvalues $\{\lambda^{(l)}_{4j}\}_{j=1}^{n^{(l)}_{+}}$
with eigenvectors $\{X^{(l)}_{4j}\}_{j=1}^{n^{(l)}_{+}}$.
We denote the number of $\lambda^{(l)}_{2j}=1$ as 
$n^{(l)}_{\downarrow}\geq0$.

We then define new fermionic operators, 
\begin{align}
\left\{
\begin{array}{ll}
\tilde{c}_{jl\uparrow}=\sum_{n=1}^{M}X^{(l)\ast}_{3jn}c_{nl\uparrow}
&(j=1,\cdots,M),\\
\tilde{c}_{jl\downarrow}^{\dagger}
=\sum_{n=1}^{M}X^{(l)}_{2jn}c_{nl\downarrow}^{\dagger}
&(j=1,\cdots,M),\\
\tilde{b}^{(l)}_{j+}=\sum_{j=1}^{n^{(l)}_{+}} X^{(l)\ast}_{4jm}b^{(l)}_{m+}
&(j=1,\cdots,n^{(l)}_{+}),\\
\tilde{b}^{(l)\dagger}_{j-}=\sum_{j=1}^{n^{(l)}_{-}}X^{(l)}_{1jm}b^{(l)\dagger}_{m-}
&(j=1,\cdots,n^{(l)}_{-}).
\end{array}
\right.
\label{eq:new_op}
\end{align}
The above transformations can be implemented by a unitary operator
${\mathcal U}^{(l)}$
which transforms the $\tilde{c}$ operators as
\begin{align}
\left\{
\begin{array}{ll}
{\mathcal U}^{(l)\dagger}\tilde{c}_{jl\uparrow}{\mathcal U}^{(l)}
=\tilde{b}^{(l)}_{j-}&(j=1,\cdots,n^{(l)}_{\uparrow}),\\
{\mathcal U}^{(l)\dagger}\tilde{c}_{n^{(l)}_{\uparrow}+j,l\uparrow}{\mathcal U}^{(l)}
=\tilde{b}^{(l)}_{n^{(l)}_{\downarrow}+j,+}&(j=1,\cdots,M-n^{(l)}_{\uparrow}),\\
{\mathcal U}^{(l)\dagger}\tilde{c}_{jl\downarrow}{\mathcal U}^{(l)}
=\tilde{b}^{(l)}_{j+}&(j=1,\cdots,n^{(l)}_{\downarrow}),\\
{\mathcal U}^{(l)\dagger}\tilde{c}_{n^{(l)}_{\downarrow}+j,l\downarrow}{\mathcal U}^{(l)}
=\tilde{b}^{(l)}_{n^{(l)}_{\uparrow}+j,-}&(j=1,\cdots,M-n^{(l)}_{\downarrow}).\\
\end{array}
\right.
\end{align}
An explicit form of the unitary operator ${\mathcal U}^{(l)}$ is found in 
Ref. \cite{pap:Lanbote1974s}.
By definition, $M\mp n^{(l)}_{\uparrow}\pm n^{(l)}_{\downarrow}=
n^{(l)}_{\pm}$, i.e. 
$\eta_l=n^{(l)}_{+}-n^{(l)}_{-}=
2(n^{(l)}_{\downarrow}-n^{(l)}_{\uparrow})=n^{(l)}$, 
are satisfied in the transformation.
The ground state wavefunction
is now obtained as 
$|{\rm GS}\rangle ={\mathcal N}\otimes_l |{\rm GS}\rangle_l$, where
\begin{align}
|{\rm GS}\rangle_l&={\mathcal U}^{(l)\dagger}|0\rangle \notag \\
&\propto \left(\prod_{j=1}^{n_{\uparrow}^{(l)}}
\tilde{c}_{j,l+\nu,\uparrow}^{\dagger}\right)
\left(\prod_{j=1}^{n_{\downarrow}^{(l)}}\tilde{c}_{j,-l,\downarrow}^{\dagger}\right)
\notag\\
&\times 
\exp \left(\sum_{j>n_{\uparrow}^{(l)}}^{M}
\sum_{j^{\prime}>n_{\downarrow}^{(l)}}^{M}
\tilde{c}_{j,l+\nu,\uparrow}^{\dagger}
F^{(l)}_{jj^{\prime}}\tilde{c}_{j^{\prime},-l,\downarrow}^{\dagger}\right)|0\rangle.
\end{align}
The coefficients $F^{(l)}_{jj'}$ are the solution of
\begin{align}
\sum_{j>n^{(l)}_{\uparrow}}^{M}
\left(X^{(l)}_{4j'}U^{(l)}_{1}X^{(l)}_{3j}\right)
F^{(l)}_{jj''}=-X^{(l)}_{4j'}U^{(l)}_{2}X^{(l)}_{2j''}
\end{align}
for $j',j''>n^{(l)}_{\downarrow}$.
It is noted that at least one of 
$n^{(l)}_{\uparrow}$ or $n^{(l)}_{\downarrow}$ must be zero
in our Hamiltonian, because the present gap functions are ``nodeless''.
In order to see this, we first divide the Hamiltonian as
$H=\sum_{ljj'}h^{(l)}_{jj'}=\sum_l
\left(
H^{(l)}_{1}+H^{(l)}_{2}+H^{(l)}_{12}\right)$,
where
\begin{align}
h_{jj'}^{(l)}&=
\left[
\begin{array}{c}
\tilde{c}^{\dagger}_{jl+\nu\uparrow}\\
\tilde{c}_{j-l\downarrow}\\
\end{array}
\right]
\tilde{H}_{{\rm BdG}jj'}^{(l)}
\left[
\begin{array}{c}
\tilde{c}_{j'l+\nu\uparrow}\\
\tilde{c}^{\dagger}_{j'-l\downarrow}\\
\end{array}
\right],\\
\tilde{H}_{{\rm BdG}}^{(l)}&=
\left[
\begin{array}{cc}
X_3^{(l)\ast}&0\\
0&X_2^{(l)}\\
\end{array}
\right]^{\dagger}
H_{{\rm BdG}}^{(l)}
\left[
\begin{array}{cc}
X_3^{(l)\ast}&0\\
0&X_2^{(l)}\\
\end{array}
\right]\notag\\
&=\left[
\begin{array}{cc}
\tilde{\varepsilon}^{(l)}&\tilde{\Delta}^{(l)}\\
\tilde{\Delta}^{(l)\dagger}&-\tilde{\varepsilon}'^{(l)}\\
\end{array}
\right].
\end{align}
Each of $H^{(l)}_{\alpha}$ ($\alpha=1,2,12$)
consists of some combinations of 
$c_{j}^{\dagger}c_{j'},c_{j}^{\dagger}c_{j'}^{\dagger}$,
and $c_{j}c_{j'}$.
$H^{(l)}_{1}$ includes the $\tilde{c}$-operators only for $j,j'=1,\cdots,
n^{(l)}_{\uparrow,\downarrow}$, 
while $H^{(l)}_{2}$ has the $\tilde{c}$-operators only for $j,j'=
n^{(l)}_{\uparrow,\downarrow},\cdots,M$.
$H^{(l)}_{12}$ connects the two sectors.
Since $|{\rm GS}\rangle_l$ is partly diagonal in the $j$-space,
the off-diagonal part vanishes $H_{12}^{(l)}|{\rm GS}_l\rangle=0$.
Therefore, $H_{1,2}|{\rm GS}\rangle_l
\propto
|{\rm GS}\rangle_l$ should be satisfied for each $i=1,2$.
Here, let us assume that both of $n_{\uparrow}^{(l)}$ and 
$n_{\downarrow}^{(l)}$ are non-zero.
Then, from $H_{1}|{\rm GS}\rangle_l\propto
|{\rm GS}\rangle_l$, we see that $\tilde{\Delta}^{(l)\dagger}_{jj'}=0$ for
$j=1,\cdots, n^{(l)}_{\downarrow}$ and $j'=1,\cdots,n^{(l)}_{\uparrow}$.
Similarly, from $H_{12}|{\rm GS}\rangle\propto
|{\rm GS}\rangle$,
it is seen that $\tilde{\Delta}^{(l)\dagger}_{jj'}=0$ for
$j=n^{(l)}_{\downarrow}+1,\cdots,M$ and $j'=1,\cdots,n^{(l)}_{\uparrow}$, and 
$\tilde{\Delta}^{(l)\dagger}_{j'j}=0$ for
$j=n^{(l)}_{\uparrow}+1,\cdots,M$ and $j'=1,\cdots,n^{(l)}_{\downarrow}$.
This means that $\tilde{\Delta}^{(l)}$ is not invertible,
and $\Delta^{(l)}=X_2^{(l)T}\tilde{\Delta}^{(l)} X_3^{(l)}$ is also
non-invertible.
Therefore, 
$\Delta^{(l)\dagger}\Delta^{(l)}$ 
must have an eigenvector with zero-eigenvalue.
Since this matrix is a representation 
of the Hermitian
operator $\Delta^{2}(\partial_x^2+\partial_y^2)^{\nu}$ 
on $L^2([0,R]\times [0,2\pi))$
with the Dirichlet boundary condition $\psi(r=R)=0$,
$\Delta^{2}(\partial_x^2+\partial_y^2)^{\nu}$
also must have a zero-eigenvale.
However, since Laplacian is positive definite on $L^2([0,R]\times [0,2\pi))$,
$\Delta^{(l)\dagger}\Delta^{(l)}$ does not have an eigenvector
with zero-eigenvalue, which means that $\Delta^{(l)}$ and 
$\tilde{\Delta}^{(l)}=X_3^{(l)T}\Delta^{(l)} X_2^{(l)}$
are invertible.
Because the resulting non-invertibility of $\tilde{\Delta}^{(l)}$ 
under the above assumption
contradicts with this,
at least one of $n^{(l)}_{\uparrow}$ or $n^{(l)}_{\downarrow}$
must be zero.
This is reasonable since it can be considered that
the factor
$\prod (\tilde{c}^{\dagger}_{l+1\uparrow}\tilde{c}^{\dagger}_{-l\downarrow})$
corresponds to nodes of gap functions,
which is seen in the ground state for nodal SFs under the periodic
boundary condition,
$|{\rm nodal}\rangle 
\propto \prod_{k:\Delta_k=0,k<k_F}c^{\dagger}_{k\uparrow}
c^{\dagger}_{-k\downarrow}
\prod_{k:\Delta_k\neq 0}\exp \left(-v_k/u_kc^{\dagger}_{k\uparrow}c^{\dagger}_{-k\downarrow}\right)
|0\rangle$,
where $(u_k, v_k)^T$ are the eigenvectors of the BdG
equation.

We can evaluate $\hat{\mathcal L}_z$ within the present scheme,
either by using the relation $n^{(l)}_{+}-n^{(l)}_{-}=
2(n^{(l)}_{\downarrow}-n^{(l)}_{\uparrow})$ 
or by directly using the explicit expression of $|{\rm GS}\rangle$.
Then, it is confirmed that Eq. (3) and Eq. (8) in the main text
are indeed 
equivalent.
This equivalence is used in order to determine $n^{(l)}_{\uparrow,\downarrow}$
in the following.

In the $p+ip$-wave SF, $\eta_l=0$ for all $l$ both in the BEC regime and 
the BCS regime. Therefore, the ground state wavefunction has
only the exponential factor
and is given by, for any $\mu$,
\begin{align}
|p{\rm GS}\rangle_l=
\exp[\tilde{c}^{\dagger}_{l+1}F^{(l)}\tilde{c}^{\dagger}_{-l}]|0\rangle.
\end{align}
For simplicity, we have suppressed the indices $j,\sigma$.
Similarly, in the BEC regimes with $\nu\geq2$,
since $\eta_l=0$ holds,
$n^{(l)}_{\uparrow}=n^{(l)}_{\downarrow}=0$ and the ground state
wavefunctions are simply of the exponetial forms.
However, in the BCS regimes with $\nu\geq2$, there exist
the factor $\prod \tilde{c}^{\dagger}_{l+\nu\uparrow}$ or
$\prod \tilde{c}^{\dagger}_{-l\downarrow}$.
Since $n_{\uparrow,\downarrow}^{(l)}\geq0$ and 
$n_{\uparrow}^{(l)}n_{\downarrow}^{(l)}=0$,
we can determine each of them from the relation
$\eta_l=n^{(l)}_+- n^{(l)}_-=2(n^{(l)}_{\downarrow}-n^{(l)}_{\uparrow})
=n^{(l)}$.
For the $d+id$-wave states, since $\eta_l=-2$ for $l_1<l<-1$,
we have $n^{(l)}_{\uparrow}=1$ and $n^{(l)}_{\downarrow}=0$ there.
Similary, since $\eta_l=+2$ for $-1<l<l_2$,
we have $n^{(l)}_{\uparrow}=0$ and $n^{(l)}_{\downarrow}=1$.
Hence, the ground state wavefunction is given by
\begin{align}
|d{\rm GS}\rangle_l=\left\{
\begin{array}{ll}
\exp[\tilde{c}^{\dagger}_{l+2}F^{(l)}\tilde{c}^{\dagger}_{-l}]
|0\rangle & (|l|>|l_{1,2}|,l=-1),\\
\tilde{c}^{\dagger}_{l+2}
\exp[\tilde{c}^{\dagger}_{l+2}F^{(l)}\tilde{c}^{\dagger}_{-l}]
|0\rangle & (l_1<l<-1),\\
\tilde{c}^{\dagger}_{-l}
\exp[\tilde{c}^{\dagger}_{l+2}F^{(l)}\tilde{c}^{\dagger}_{-l}]
|0\rangle & (-1<l<l_2).
\end{array}
\right.
\label{eq:dGS}
\end{align}
We note that, $c_{n,l+2} (c_{n,-l})$-fermions 
have lower energies than
$c_{n,-l} (c_{n,l+2})$-fermions with the same radial quantum number
$n$ for $-l_1<l<-1 (-1<l<l_2)$.
This would be the reason why the
$\tilde{c}_{l+2} (\tilde{c}_{-l})$-fermions are unpaired 
for $-l_1<l<-1 (-1<l<l_2)$, but not vice versa.
Similar wavefunctions with unpaired fermions represented by
extra $\tilde{c}^{\dagger}_{l+\nu},
\tilde{c}^{\dagger}_{-l}$ operators are also obtained for $\nu\geq3$
in the BCS regimes.

\end{document}